\documentclass[iop, twocolumn]{aastex63}

\usepackage{lineno}
\usepackage{graphicx}

\usepackage{graphicx}
\usepackage{float}
\usepackage{gensymb}

\received{XXX}
\revised{XXX}
\accepted{XXX}
\submitjournal{XXXX}

\newcommand{\dmunit}{pc\,cm$^{-3}$}

\shortauthors{Xu et al.}

\begin{document}
%\linenumbers
\title{Discovery of \textbf{15} new pulsars at high Galactic Latitudes with FAST} 

\correspondingauthor{Shi Dai}\email{shi.dai@csiro.au}
\correspondingauthor{Qijun Zhi}\email{qjzhi@gznu.edu.cn}

%\nocollaboration{1}
\author[0009-0006-3224-4319]{Xin Xu}
\affiliation{School of Mathematical Science, Guizhou Normal University, Guiyang 550001, China}
\affiliation{Guizhou Provincial Key Laboratory of Radio Astronomy and Data Processing, Guizhou Normal University, Guiyang 550001, People's Republic of China.}

\author[0000-0002-9618-2499]{Shi Dai}
\affiliation{Australia Telescope National Facility, CSIRO, Space and Astronomy, PO Box 76, Epping, NSW 1710, Australia}
\affiliation{Western Sydney University, Locked Bag 1797, Penrith South DC, NSW 2751, Australia}

\author[0000-0001-9389-5197]{Qijun Zhi}
\affiliation{Guizhou Provincial Key Laboratory of Radio Astronomy and Data Processing, Guizhou Normal University, Guiyang 550001, People's Republic of China.}
\affiliation{School of Physics and Electronic Science, Guizhou Normal University, Guiyang 550001, People's Republic of China}

\author[0000-0002-1052-1120]{Juntao Bai}
\affiliation{Xinjiang Astronomical Observatory, Chinese Academy of Sciences, Urumqi, Xinjiang 830011, People's Republic of China}
\affiliation{School of Astronomy and Space Science, University of Chinese Academy of Sciences, Beijing, 100049, People's Republic of China}

\author[0000-0003-4962-145X]{Joanna Berteaud}
\affiliation{University of Maryland, Department of Astronomy, College Park, MD, 20742 USA}
\affiliation{NASA Goddard Space Flight Center, Code 662, Greenbelt, MD, 20771 USA}

\author[0000-0001-7722-6145]{Francesca Calore}
\affiliation{LAPTh, CNRS, USMB, 74940 Annecy, France}

\author[0000-0003-0724-2742]{Ma\text{\"i}ca Clavel}
\affiliation{Univ. Grenoble Alpes, CNRS, IPAG, 38000 Grenoble, France}

\author[0000-0001-5105-4058]{Weiwei Zhu}
\affiliation{National Astronomical Observatories, Chinese Academy of Sciences, Beijing 100101, China}

\author[0000-0003-3010-7661]{Di Li}
\affiliation{Department of Astronomy, Tsinghua University, Beijing 100190, China}
\affiliation{National Astronomical Observatories, Chinese Academy of Sciences, Beijing 100101, China}

\author[0000-0001-9389-5197]{Rushuang Zhao}
\affiliation{Guizhou Provincial Key Laboratory of Radio Astronomy and Data Processing, Guizhou Normal University, Guiyang 550001, People's Republic of China.}
\affiliation{School of Physics and Electronic Science, Guizhou Normal University, Guiyang 550001, People's Republic of China}

\author[0000-0002-9042-3044]{Renxin Xu}
\affiliation{School of Physics and State Key Laboratory of Nuclear Physics and Technology, Peking University, Beijing 100871, China}

\author[0000-0002-9042-3044]{Guojun Qiao}
\affiliation{School of Physics and State Key Laboratory of Nuclear Physics and Technology, Peking University, Beijing 100871, China}

%
%\nocollaboration{12}

\begin{abstract}
We present the discovery and timing results of 15 pulsars discovered in a high Galactic latitude survey conducted with the Five-hundred-meter Aperture Spherical Telescope (FAST). The survey targeted a region as close as possible to the Galactic Center, encompassing an area near the Galactic Bulge.
The newly discovered pulsars consist of eleven normal pulsars and four millisecond pulsars (MSPs). Among the MSPs, three are identified in binary systems with orbital periods of $\sim3.1$, 4.6 and 12.5\,days, respectively. We have successfully obtained coherent timing solutions for three of the normal pulsars (PSRs~J1745$-$0059, J1746$-$0156 and J1800$-$0059). Furthermore, within our data set we found that four pulsars (three new and one known) show mode-changing and/or subpulse drifting phenomena. Comparing our discoveries with simulations of the Galactic disk and Bulge MSP populations indicates that these new pulsars are most likely located in the disk. 
Nonetheless, our discoveries demonstrate that deep surveys at high Galactic latitudes have significant potential to enhance our understanding of the MSP population in the direction of the Bulge.
\end{abstract}

\keywords{pulsars: general}

%%%%%%%%%%%%%%%%%%%%%%%%%%%%%%%
\section{Introduction}
\label{sec_1}

To date, nearly 4,000 pulsars have been discovered~\citep[e.g.,][]{2005AJ....129.1993Manchester}, with the discovery rate accelerating in recent years largely due to deep surveys conducted by the Five-hundred-meter Aperture Spherical Telescope~\citep[FAST;][]{2018IMMag..19..112LiD, 2020MNRAS.495.3515C, 2021RAA....21..107HanJL, 2021MNRAS.508..300C, 2021ApJ...915L..28PanZC, 2023RAA....23j4001Z, 2023MNRAS.526.2645S,  2023ApJS..269...56Z, 2024ApJ...975...88Z, 2024ApJ...969L...7YinDJ} and the MeerKAT telescope~\citep{2022A&A...664A..27Ridolfi, 2022MNRAS.513.1386V, 2023MNRAS.520.3847C, 2023MNRAS.524.1291P, 2024MNRAS.531.2835Carli, 2024MNRAS.533.2570P,2024MNRAS.531.3579Turner,2024A&A...686A.166P}. \citet{2024ApJ...960...79ZhiQJ} conducted a pilot survey at intermediate Galactic latitudes, demonstrating that FAST surveys in these regions have the potential to uncover hundreds of millisecond pulsars (MSPs). 
Most of these deep pulsar surveys have concentrated on the Galactic plane or globular clusters. At high Galactic latitudes, previous pulsar surveys have primarily utilized low radio frequencies \citep{2014A&A...570A..60Coenen, 2019A&A...626A.104Sanidas, 2023PASA...40...21Bhat, 2023PASA...40...20B}, drift scans \citep{2000ApJ...545.1007Lommen, 2013ApJ...763...80Boyles, 2013ApJ...763...81Lynch, 2018IMMag..19..112LiD, 2022MNRAS.509.1929Perera}, or short integration times per pointing \citep{2006MNRAS.368..283Burgay, 2010MNRAS.409..619K, 2018MNRAS.473..116Keane}, making them generally sensitive only to bright or nearby pulsars compared to Galactic plane surveys. This limitation constrains our understanding of the pulsar population in these regions, particularly towards areas such as the Galactic Bulge~\citep[e.g.,][]{cdd+16}.

The dense stellar environment of the Galactic center and Bulge is expected to host a substantial population of MSPs \citep{cdd+16,2022NatAs...6..703G}. This population has been proposed as the origin of the enigmatic Galactic Center $\gamma$-ray excess~\citep[GCE;][]{2020ARNPS..70..455M}, although whether dark matter annihilation contributes to the observed diffuse $\gamma$-ray emission remains an open question~\citep[e.g.,][]{2015PhRvD..91f3003C,2018NatAs...2..387M,2021PhRvL.127p1102C,2024MNRAS.530.4395S}. Recent studies combining $\gamma$-ray and X-ray observations of the Galactic Bulge suggest that many unidentified X-ray sources in the region are consistent with being MSPs that have yet to be identified~\citep{2021PhRvD.104d3007B}.
\citet{cdd+16} conducted simulations of the MSP population in the Galactic Bulge, demonstrating that their distribution could extend up to tens of degrees above (and below) the Galactic plane, despite their distribution being strongly peaked towards the Galactic center. Given the considerable distance to the Galactic center, probing this population requires deep surveys that cover a substantial portion of the Bulge. The presence or absence of detectable Bulge MSPs would have profound implications for understanding the origin of the GCE and could contribute to refining dark matter theories.

If such a population exists, determining whether Bulge MSPs share the same formation and evolutionary history as MSPs in the Galactic disk and globular clusters would provide valuable insights into the stellar evolution and dynamics of the Bulge. Discovering these MSPs would enable direct comparisons of their ages, magnetic field strengths, and binary system properties (e.g., binary fraction, orbital periods) with those of disk and globular cluster MSPs. Additionally, finding pulsars in the Bulge would offer new opportunities to study the distribution of free electrons~\citep[e.g.,][]{2017ApJ...835...29YaoJM} and the magnetic field structure \citep{2012MNRAS.427..664Schnitzeler} in the inner regions of our Galaxy.

In this paper, we present the results of a high Galactic latitude pulsar survey conducted with FAST. The survey area was selected to be as close to the Galactic Bulge as possible, constrained by the sky coverage accessible with FAST's snapshot observing mode~\citep{2021RAA....21..107HanJL}. Thanks to FAST's exceptional sensitivity, this represents the deepest survey at mid- and high-latitudes, skimming the outskirts of the Galactic Bulge.
%this represents the deepest survey of the targeted Bulge region to date. 
Section~\ref{sec_2} describes the survey and associated timing campaign, while Section~\ref{sec_3} presents the results. Finally, conclusions and a discussion are provided in Section~\ref{sec_4}.

%%%%%%%%%%%%%%%%%%%%%%%%%%%%%%%%%%%%%%%%%%
%%%%%%%%%%%%%%%%%%%%%%%%%%%%%%%%%%%%%%%%%%
\section{Observations and Data Reduction} \label{sec_2}

\subsection{FAST survey and pulsar search}

\begin{figure}
  \centering 
  \includegraphics[width=0.5\textwidth]{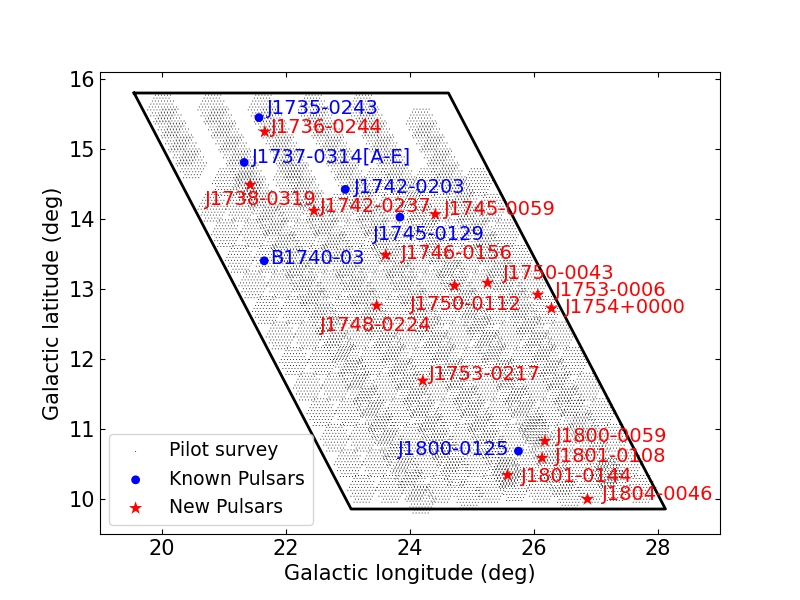} 
  \caption{ A plot of the sky coverage map of this survey in galactic coordinates. Black dots represent each survey beam positions, blue dots mark the positions of known pulsars within the survey region, and red stars indicate the locations of newly discovered pulsars. The area enclosed by the solid black line represents the region we plan to survey.
    \label{fig_skyglgb}}
\end{figure}

The 19-beam L-band focal plane array of FAST \citep{2018IMMag..19..112LiD} was used to survey a region at high Galactic latitude. The observing band covered a frequency range of 1.0 to 1.5\,GHz \citep{2020RAA....20...64JiangP}. A total of 238 observations were carried out, utilizing the FAST snapshot observing mode and covering Galactic longitudes and latitudes of $21\degree <l< 27\degree$ and $10\degree<b<16\degree$. Each snapshot observation consists of four independent pointings, offset from one another, to ensure uniform sky coverage while minimizing slewing time. More details on the snapshot mode can be found in \citet{2021RAA....21..107HanJL}. The integration time per pointing is five minutes. 
In Fig.\,\ref{fig_skyglgb}, we show the footprint of each pointing as grey regions. The lower boundary in Galactic longitude of our survey is set by the limiting declination of FAST in the snapshot mode.    
The FAST ROACH backend was used in its pulsar-search mode, with 4096 channels across 500 MHz of bandwidth and 49\,$\rm \mu$s sampling rate. Total intensity was recorded with 8-bit sampling.

A periodicity search was conducted using the pulsar searching software package \textit{PRESTO} \citep{2001PhDT.......123Ransom}, covering a dispersion measure (DM) range of $\rm 0-1000\,pc\,cm^{-3}$. Radio frequency interference (RFI) was masked using the \textit{rfifind} routine. To account for potential orbital modulation of pulsar periodic signals, we searched for signals drifting by up to $\pm200/n_{\rm h}$ bins in the Fourier domain by setting $z_{\rm max}=200$ \citep{2002AJ....124.1788Ransom}, where $n_{h}$ is the largest harmonic at which a signal is detected (up to 8 harmonics were summed). Additionally, single pulse candidates with a signal-to-noise ratio (S/N) greater than eight were identified using the \textit{single\_pulse\_search.py} routine. This routine was applied to each de-dispersed time series and boxcar filtering parameters, using filter widths ranging from 1 to 300 samples. Burst candidates were manually examined after removing narrowband and impulsive RFI.

Through the methods described above, we discovered 15 new pulsars, all of which have been successfully confirmed by follow-up observations. In Fig.\,\ref{fig_skyglgb}, blue circles represent the position of known pulsars within the survey region, and red stars show the locations of newly discovered pulsars. The basic parameters of our discoveries are presented in Table\,\ref{tab_allpsrpara}. A detailed description of these new pulsars is provided in the next section.
Further search of this data set is currently ongoing and updates will be continuously provided on our project webpage\,\footnote{https://astrolab.gznu.edu.cn/kxyj.htm}.

We detected single pulses from newly discovered pulsars J1736$-$0245, J1745$-$0059, J1746$-$0156, J1800$-$0059 and J1801$-$0108, and known pulsars B1740$-$03, J1735$-$0243, J1745$-$0129 and J1800$-$0125. 
Studies of their single pulses are presented in Section~\ref{sec_3.3}. No rotating radio transient sources (RRATs) or fast radio bursts (FRBs) were detected in this survey. Assuming a system equivalent flux density of 1.25\,Jy \citep{2020RAA....20...64JiangP}, our sensitivity of detecting single pulses with a width of 5\,ms and a S/N of 8 is 4.5\,mJy.

%%%%%%%%%%%%%%%%%%%%%%%%%%%%%%%%%%%%%%%%%%%%%%%%

\subsection{Follow-up timing observations}

Three pulsars underwent regular timing observations, namely J1745$-$0059, J1746$-$0156, and J1800$-$0059. Among these, J1745$-$0059 and J1800$-$0059, two of the brightest pulsars in our sample, were observed with Murriyang, CSIRO's Parkes 64-meter telescope. 
The Ultra-Wideband Low (UWL) system\,\citep{2020PASA...37...12Hobbs} was employed in coherently de-dispersed search mode, recording data with 2-bit sampling every 64\,$\mu$s in each of the 1\,MHz wide frequency channels covering a total bandwidth of 3328\,MHz between 704 and 4032\,MHz. Only total intensity data were recorded. The integration times of J1745$-$0059 and J1800$-$0059 ranges from 10 to 30 minutes and 45 to 80 minutes per observation, respectively.
PSR~J1746$-$0156 was observed and timed with FAST using the central beam of the 19-beam receiver. Data were recorded in pulsar search mode with configurations identical to those used in the survey, including full polarization information. The integration time of this pulsar ranges from 180 to 300 seconds per observation.

Furthermore, we carried out eight follow-up observations of three MSPs (J1742$-$0237, J1748$-$0224, and J1801$-$0144) in binary systems with FAST, aiming to obtain their basic orbital parameters. The data recording and configuration were identical to those used for J1746$-$0156, with each pulsar having an integration time of five minutes. Additionally, we performed six observations of J1801$-$0144 using Murriyang to refine its orbital parameters. %The other two MSPs cannot be detected by Murriyang. 
The other two MSPs are too faint to be detected by Murriyang with an integration time of $\sim1$\,hr and were therefore not followed up with Murriyang.

To obtain a coherent timing solution, search mode observations were folded with the apparent spin period of each pulsar determined at each observing epoch using the {\sc DSPSR} software package \citep{2011PASA...28....1Van_Straten}, with a subintegration length of 30 seconds. 
Narrowband and impulsive RFI were manually removed from each subintegration. 
Then, each observation was averaged in time to create subintegrations with a length of a few minutes, and pulse times of arrival (ToAs) were measured for each subintegration using the {\sc pat} routine of the {\sc PSRCHIVE} software package \citep{2012AR&T....9..237Van_Straten}. 
Finally, the timing analysis was performed for each pulsar using the {\sc TEMPO2} software package \citep{2006MNRAS.369..655Hobbs}, where the timing analysis was performed using the Barycentric Coordinate Time (TCB), TT(TAI) clock standard, and JPL DE438 solar system ephemeris.

We successfully obtained coherent timing solutions for three pulsars (J1745$-$0059, J1746$-$0156, J1800$-$0059). For these pulsars, we re-folded the search mode data and averaged the time and frequency of each observation to generate high S/N pulse profiles.
ToAs were measured using these high S/N profiles and we repeated our timing analysis to measure their spin and astrometric parameters. Throughout our timing analysis, {\sc TEMPO2} fitting with ToA errors (known as `MODE 1') was used and the weighted root-mean-square (Wrms) of timing residuals were reported in Fig.\,\ref{fig_res} and Table\,\ref{tab_timing}. To refine our DM measurements, for each pulsar we selected a high S/N observation and divided the bandwidth into four frequency subbands. We then measured a ToA from each subband and fitted for the DM using {\sc TEMPO2}. Our timing results will be presented in the next Section.

\begin{figure*}
  \centering \includegraphics[width=0.95\textwidth]{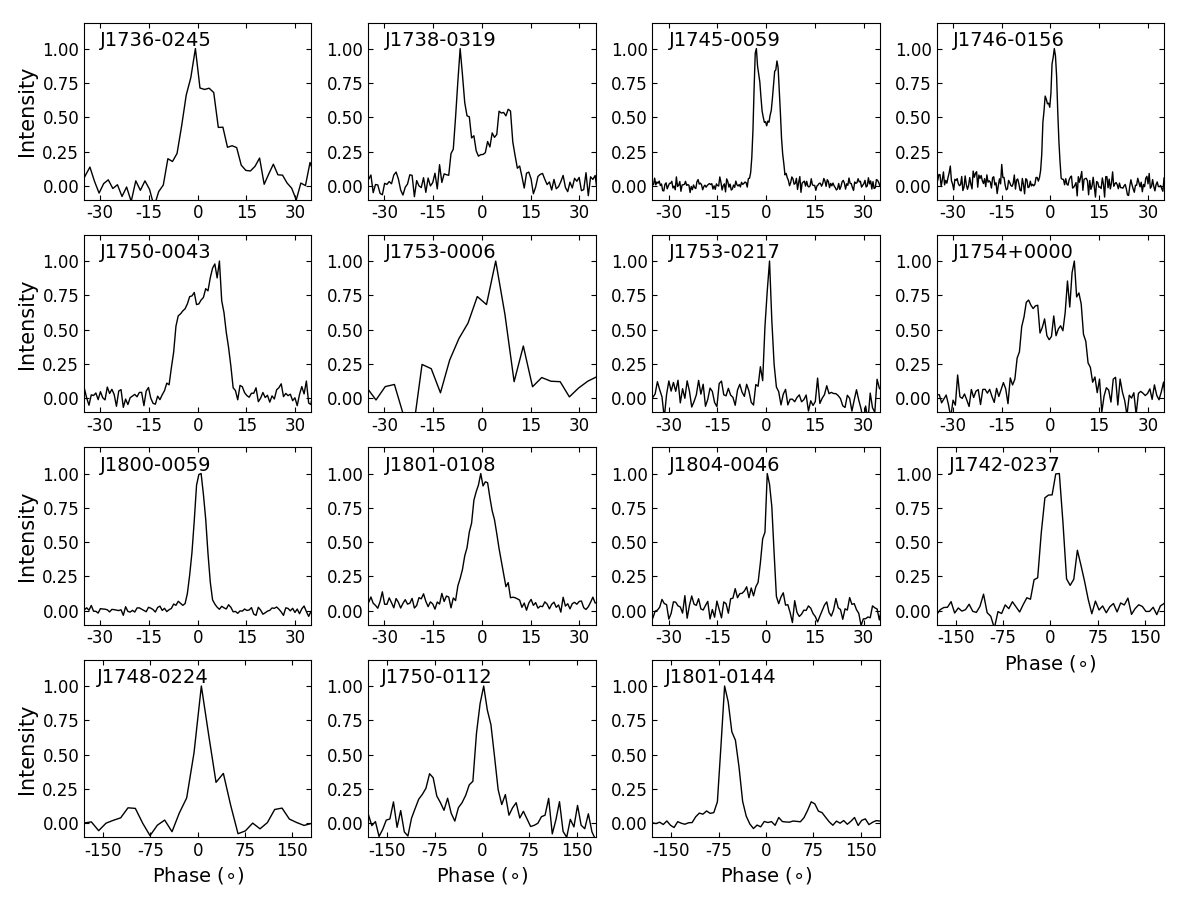}   
  \caption{Integrated pulse profiles of the 15 new pulsars were obtained with FAST at the center frequency of 1250~MHz. The first eleven panels display the 5-minute integrated pulse profiles of the eleven normal pulsars, while the next four panels present the 20-minute integrated pulse profiles of the four MSPs. All pulse profiles have been normalized.
    \label{fig_allprofiles}}
\end{figure*}

\begin{figure*}
  \centering 
  \includegraphics[width=5.5cm,height=5.3cm]{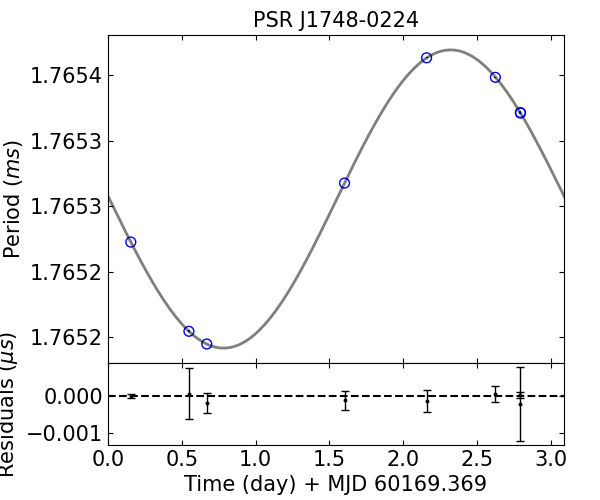}
  \includegraphics[width=5.5cm,height=5.3cm]{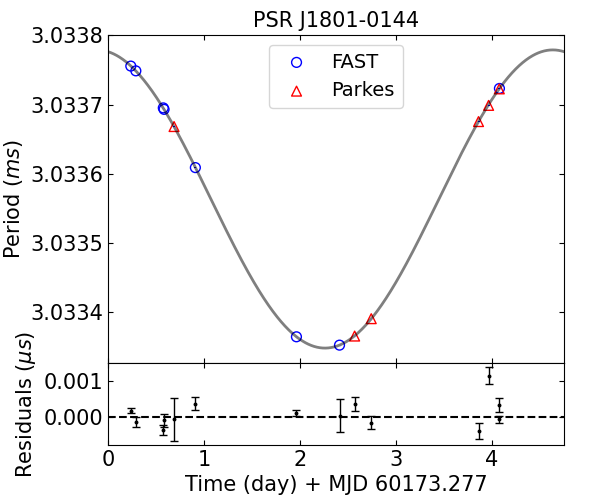}     
  \includegraphics[width=5.5cm,height=5.3cm]{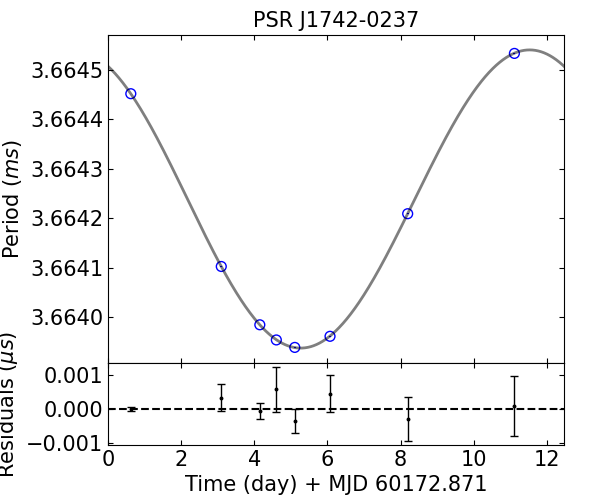}
  \caption{Variations in the observation period of three pulsars in binary systems were recorded using FAST and Murriyang. Each data point corresponds to a five minute (FAST) or one hour (Murriyang) observation plotted against a single orbital period. The black curves depict orbital fits of the period changes, while the lower panel shows the residuals of the fit.
    \label{fig_orbitfit}}
\end{figure*}

%%%%%%%%%%%%%%%%%%%%%%%%%%%%%%%%%%%%%%%%%%
%%%%%%%%%%%%%%%%%%%%%%%%%%%%%%%%%%%%%%%%%%

\section{Results}\label{sec_3}
We discovered 15 new pulsars, including eleven normal pulsars and four MSPs, with a total observing time of 118.5 hours covering approximately $\sim 27$ square degrees (see Table\,\ref{tab_allpsrpara}).
Additionally, we detected six previously known pulsars in the surveyed area: B1740$-$03~\citep{1978MNRAS.185..409Manchester}, J1735$-$0243~\citep{2013ApJ...763...80Boyles}, J1737$-$0314A~\citep{2021ApJ...915L..28PanZC}, J1742$-$0203~\citep{2020ApJ...892...76McEwen}, J1745$-$0129~\citep{2001MNRAS.326..358Edwards}, J1800$-$0125~\citep{2006ApJ...652.1499Crawford}, with J1737$-$0314A being a pulsar in the globular cluster NGC~6402~\citep{2021ApJ...915L..28PanZC}. This globular cluster hosts six MSPs, but due to our limited sensitivity (with a five-minute integration time per pointing), we could not detect the remaining five pulsars. This section will provide a detailed discussion of all newly discovered pulsars and pulsars with special single-pulse phenomena.

\begin{table*}
\begin{center}
\caption{The basic parameters of all 15 new pulsars, where three of the pulsars for which coherent timing solutions have been obtained are indicated by asterisks. For three MSPs in binary systems, their orbital parameters were determined by fitting a circular orbit model to the variation of their barycentric periods in time. The minimum, median, and maximum companion masses for binary systems were estimated assuming a pulsar mass of 1.35 $M_{\odot}$ and an inclination angle of 90, 60, and 26 degrees, respectively. Pulsar distances were estimated using free electron density models of \citet{2002astro.ph..7156C} and \citet{YMW16}.}
\label{tab_allpsrpara}
\begin{tabular}{cccccccccccc}
\hline
\hline
Pulsar & RA & DEC & Period & DM & \multicolumn{2}{c}{Dist} & $P_b$ & $\chi_p$ & Companion mass & Type\\ 
  & J2000 & J2000 & (s) & (\dmunit) & \multicolumn{2}{c}{(kpc)} & (days)  & ($ls$) & (M$_{\odot}$) & \\
    &  &  &  &  & YMW16 & NE2001 &  &  &  & \\

\hline
\multicolumn{11}{c}{Normal Pulsars}\\
\hline
J1736$-$0245 & 17:36:41 & $-$02:45 & 0.139298 & 44.3(1) & 0.7 & 1.6 & $-$ & $-$ & $-$ & Isolated \\
J1738$-$0319 & 17:38:55 & $-$03:19 & 1.369302 & 85.0(4) & 2.3 & 2.9 & $-$ & $-$ & $-$ & Isolated \\
J1745$-$0059$^{*}$ & 17:45:56.70 & $-$00:58:54.0 & 0.679457 & 68.63(3) & 3.3 & 2.4 & $-$ & $-$ & $-$ & Isolated \\
J1746$-$0156$^{*}$ & 17:46:30.93 & $-$01:56:31 & 1.829039 & 92.94(8) & 5.2 & 3.2 & $-$ & $-$ & $-$ & Isolated \\
J1750$-$0043 & 17:50:57 & $-$00:43 & 0.571810 & 85.1(4) & 5.3 & 2.9 & $-$ & $-$ & $-$ & Isolated \\
J1753$-$0006 & 17:53:02 & $-$00:06 & 0.263349 & 101(1) & 8.9 & 3.6 & $-$ & $-$ & $-$ & Isolated \\
J1753$-$0217 & 17:53:58 & $-$02:17 & 1.039570 & 91.5(5) &  4.5 & 3.0  & $-$ & $-$ & $-$ & Isolated \\
J1754+0000 & 17:54:07 & +00:00 & 0.0611325 & 104.71(5) & 10.0 & 3.7 & $-$ & $-$ & $-$ & Isolated \\
J1800$-$0059$^{*}$ & 18:00:39.07 & $-$00:59:13.6 & 1.192542 & 84.2(2) & 4.5 & 2.7 & $-$ & $-$ & $-$ & Isolated \\
J1801$-$0108 & 18:01:25 & $-$01:08 & 1.072109 & 86.3(4) & 4.6 & 2.7 & $-$ & $-$ & $-$ & Isolated \\
J1804$-$0046 & 18:04:51 & $-$00:46 & 0.635152 & 86.8(2) & 4.7 & 2.7 & $-$ & $-$ & $-$ & Isolated \\
\hline
\multicolumn{11}{c}{Millisecond pulsars}\\
\hline
J1742$-$0237 & 17:42:08 & $-$02:37 & 0.00366367 & 65.685(6) & 1.5 & 2.2 & 12.48 & 14.14 & $0.39 < 0.46 < 1.12$ & Binary \\
J1748$-$0224 & 17:48:48 & $-$02:24 & 0.00176509 & 75.899(3) & 2.8 & 2.5 & 3.09 & 2.74 & $0.18 < 0.21 < 0.45$ & Binary \\
J1750$-$0112 & 17:50:07 & $-$01:12 & 0.00889597 & 72.97(3) & 3.5 & 2.5 & $-$ & $-$ & $-$ & Isolated \\
J1801$-$0144 & 18:01:15 & $-$01:44 & 0.00303345 & 60.021(2) & 2.1 & 2.0 & 4.76 & 4.64 & $0.23 < 0.27 < 0.60$ & Binary \\
\hline
\end{tabular}
\end{center}
\end{table*}

%%%%%%%%%%%%%

\subsection{Normal pulsars}

The newly discovered pulsars include eleven isolated normal pulsars. Their basic parameters are listed in Table\,\ref{tab_allpsrpara}, including their names, coordinates (right ascension and declination), spin periods, and DM. One of them, J1754+0000, has a spin period of $\sim 0.06\,\rm s$, making it potentially a young pulsar. Measurements of its spin-down rate through follow-up timing observations are required to understand its nature.

We estimated the distances to these pulsars using free electron density models of YMW16~\citep{2017ApJ...835...29YaoJM} and NE2001~\citep{2002astro.ph..7156C}. The distances are provided in the sixth and seventh columns of Table\,\ref{tab_allpsrpara}. For several pulsars, the distances estimated by the two models differ significantly, as seen in cases like PSRs~J1754+0000 and J1753$-$0006, where the YMW16 model predicts substantially greater distances compared to the NE2001 model.

The integrated pulse profiles of these normal pulsars observed with FAST are shown in the first eleven panels of Figure\,\ref{fig_allprofiles}, using a five-minute integration at a central frequency of 1250~MHz. 
The results indicate that the integrated profiles of five pulsars (PSRs~J1753$-$0006, J1800$-$0059, J1801$-$0108, J1753$-$0217 and J1736$-$0244) exhibit a single-component pulse shape, with PSR~J1753$-$0217 having the narrowest radiation window of just a few degrees. 
In contrast, the integrated pulse profile of J1753$-$0006 and J1736$-$0244 are asymmetric, with the much steeper right side of the former and the opposite of the latter, suggesting the presence of multiple underlying components that require verification through longer observations. 
The other six pulsars, PSRs~J1754+0000, J1750$-$0043, J1746$-$0156, J1804$-$0046, J1745$-$0059 and J1738$-$0319, display relatively complex integrated pulse profiles, each consisting of at least two components. J1746$-$0156's integrated profile features two closely separated components, while J1754+0000, J1745$-$0059 and J1738$-$0319 show two distinctly separated components with a clear bridge of radiation between them. However, J1738$-$0319 appears to have a multi-component profile. 
Furthermore, the pulse profile of J1804$-$0046 shows a weak component at $-8^{\circ}$, with a peak intensity that is only one-fifth of the overall integral profile. 
It's worth noting that the five minute integration with FAST does not provide an integrated pulse profile with a high signal-to-noise ratio. The shape of the integrated pulse profiles of these pulsars may change under longer observations.

\begin{figure}[hp]
  \centering \includegraphics[width=0.45\textwidth]{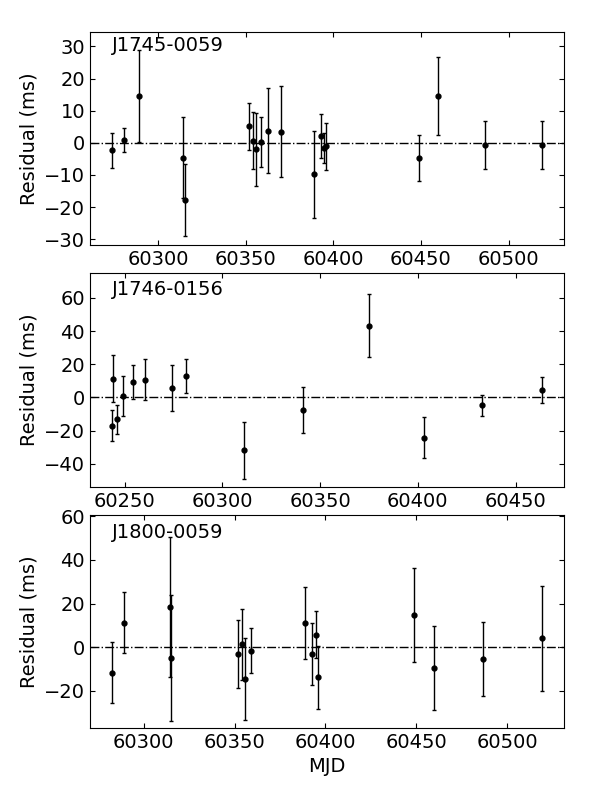}   
  \caption{The time residuals of the three normal pulsars are plotted as a function of MJD. The weighted root-mean-square of timing residuals for each pulsar are listed in Table\,\ref{tab_timing}.  The timing of J1746$-$0156 was carried out with FAST, while J1745$-$0059 and J1800$-$0059 were observed with Murriyang.
    \label{fig_res}}
\end{figure}

To date, we have successfully obtained coherent timing solutions for three (J1745$-$0059, J1746$-$0156 and J1800$-$0059) of the newly discovered pulsars. The best-fit timing parameters for these pulsars are presented in Table\,\ref{tab_timing}, while the corresponding timing residuals are shown in Figure\,\ref{fig_res}. 
For other normal pulsars where coherent timing solutions have not yet been achieved, results will be reported in a future publication. Notably, we carried out three one-hour observations of J1801$-$0108 using Murriyang, the Parkes telescope, but no pulsar signals were detected. Consequently, follow-up timing studies for this pulsar will need to rely on high-sensitivity telescopes like FAST.

%%%%%%%%%%%%%%%%%%%%%%%%%%%%%%

\subsection{Millisecond pulsars}

So far, we have discovered four new MSPs, all with spin periods of less than 10\,ms. Except for PSR J1750$-$0112, the spin periods of the other three MSPs (PSRs J1742$-$023, J1748$-$0224, and J1801$-$0144) exhibit significant variations in each observation, indicating that they are in binary systems. 
For those three pulsars in binary systems, we measured their orbital parameters by fitting the variation of their barycentric periods as a function of time obtained from FAST and Murriyang observations using the \textit{fitorbit}\,\footnote{https://github.com/vivekvenkris/fitorbit} program. The fitting results indicate that all three pulsars are in nearly circular orbits with small eccentricities ($e<10^{-3}$) .
Figure\,\ref{fig_orbitfit} shows the fitting results for their period changes and the corresponding residuals, where blue circles indicate FAST observations and red triangles indicate Murriyang observations.
The binary orbital period ($P_b$) and the projection of the semi-major axis along the line of sight ($\chi_p$) for these pulsars are listed in the eighth and ninth columns of Table\,\ref{tab_allpsrpara}, respectively. Additional timing observations are currently underway to obtain coherent timing solutions and determine their orbital parameters.

PSR~J1750$-$0112 is an MSP with a spin period of $\sim\,8.896\, \rm 
ms$. Three observations with FAST showed negligible changes in its observed period, within the limits of uncertainty. Therefore, it is likely an isolated MSP. Using the NE2001 and YMW16 models, we estimated the distance to this pulsar, finding it to be the most distant of the four new MSPs. 
Its integrated pulse profile, shown in Figure\,\ref{fig_allprofiles}, is based on twenty minutes of FAST observation. The profile displays two obvious components, with a wide pulse emission window covering almost half of the entire pulse period. 
The separation between the two components is $\sim 85 ^{\circ}$. Due to the low S/N ratio, the presence of bridge radiation between the components remains uncertain and requires verification through longer observations.

For the three pulsars in binary systems, J1748$-$0224, J1801$-$0144, and J1742$-$0237, preliminary orbital solutions suggest orbital periods of approximately 3.09, 4.76, and 12.48 days, respectively. Meanwhile, these pulsars all have very short spin periods, with J1748$-$0224 having the shortest at 1.765~ms. 
To estimate their companion masses, we employed the \citet{2004hpa..book.....L} method, assuming the pulsar mass to be $M_{\rm p}=1.35$ \, M$_{\odot}$, and the minimum, median, and maximum companion masses were estimated with an inclination angle of $i=90^{\circ}$, $60^{\circ}$ and $26^{\circ}$, respectively. 
The minimum, median, and maximum companion mass of these three pulsars are listed in the tenth column of Table\,\ref{tab_allpsrpara}.
The results suggest that all three MSPs have white dwarf (WD) as their companion. Among them, J1742$-$0237 has the largest companion mass, while J1748$-$0224 has the smallest.

We obtained the integrated pulse profiles of these three pulsars using FAST observations with twenty minutes of integration at a central frequency of 1250 MHz. The results are shown in the last three panels of Figure\,\ref{fig_allprofiles}. 
For J1748$-$0224, the integrated pulse profile is single-component pulse shape. 
For J1801$-$0144, the integrated pulse profile shows both the main pulse (MP) and the interpulse (IP), with a separation of $\sim\,150^{\circ}$. The peak intensity of the IP is only one-fifth that of the MP. Additionally, the MP has a weak leading component. We did not detect any bridge radiation between the MP and IP. The integrated profile of J1742$-$0237 exhibits a distinct tail component.

\begin{table*}
\renewcommand\arraystretch{1}
\begin{center}
\caption{Parameters for three normal pulsars with timing solutions, including right ascension, declination, rotation period, and derived parameters.}
\label{tab_timing}
\begin{tabular}{lccc}
\hline
\hline
%\multicolumn{4}{c}{Pulsars with timing solutions}\\
 & J1745$-$0059 & J1746$-$0156  & J1800$-$0059 \\
\hline
RAJ(J2000) & 17:45:56.70(1) & 17:46:30.93(4) & 18:00:39.07(3) \\
DECJ(J2000) & $-$00:58:54.0(1)  & $-$01:56:31(1)  & $-$00:59:13.6(3) \\
$\nu$ (Hz) & 1.47163492631(2) & 0.54673292488(3) & 0.83854355767(3) \\
$\dot{\nu}$ (Hz/s) & $-9.4(2) \times 10^{-16}$ & $-5.3(3)\times 10^{-16}$ & $-1.9(2)\times 10^{-17}$ \\
PEPOCH (MJD) & 60396.3 & 60353.5 & 60400.8 \\
Time span (MJD) & 60273-60519 & 60243-60463 & 60282-60519 \\
DM ($\rm cm^{-3}\,pc$) & 68.63(3) & 92.94(8) & 84.2(2) \\
%RM (rad\,m$^{-2}$) & - & 598(9) & - \\
$W_{rms}$(ms) & 0.085 & 0.21 & 0.14 \\
\hline
\multicolumn{4}{c}{Derived parameters}\\
\hline
GL (degree) & 24.41 & 23.61 & 26.18 \\
GB (degree) & 14.07 & 13.48 & 10.82 \\
$P$ (s) & 0.679516354310(8) & 1.82904660482(9) & 1.19254389453(4) \\
$\dot{P}$ (s\,s$^{-1}$) & 4.36(7)$\times 10^{-16}$ & 1.77(8)$\times 10^{-15}$ & 2.8(3)$\times 10^{-16}$\\
$\tau_c$ (Myr) & 24.73 & 16.401 & 68.33 \\
$B_s$ (G) & 5.51$\times 10^{11}$ & 1.82$\times 10^{12}$ & 5.81$\times 10^{11}$\\
\hline
\end{tabular}
\end{center}
\end{table*}

%%%%%%%%%%%%%%%%%%%%%%%

\subsection{Single pulse phenomenon}
\label{sec_3.3}

\begin{figure*}
  \centering 
  \includegraphics[width=4cm,height=5.5cm]{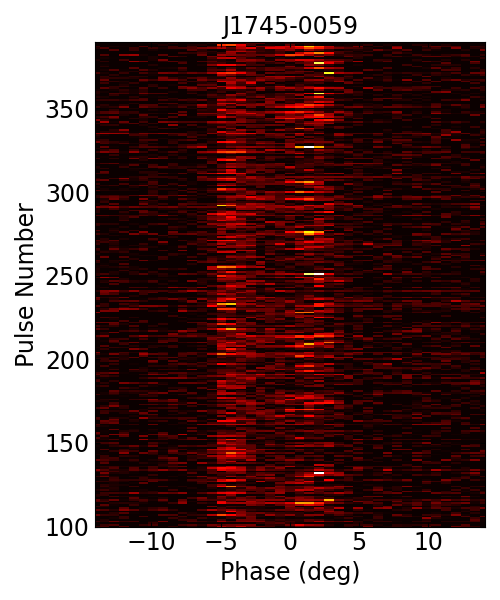}
  \includegraphics[width=4cm,height=5.5cm]{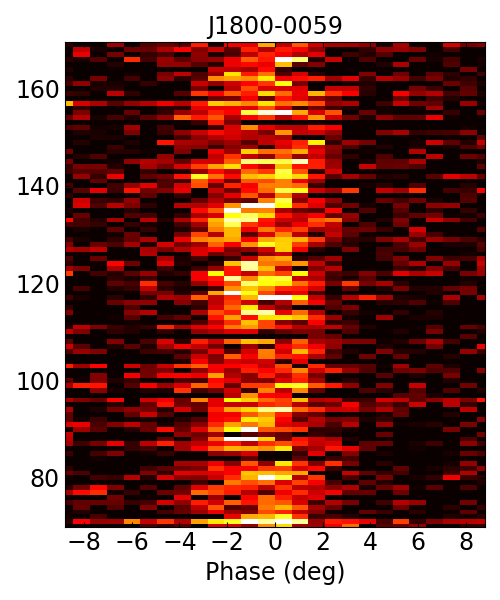}
  \includegraphics[width=4cm,height=5.5cm]{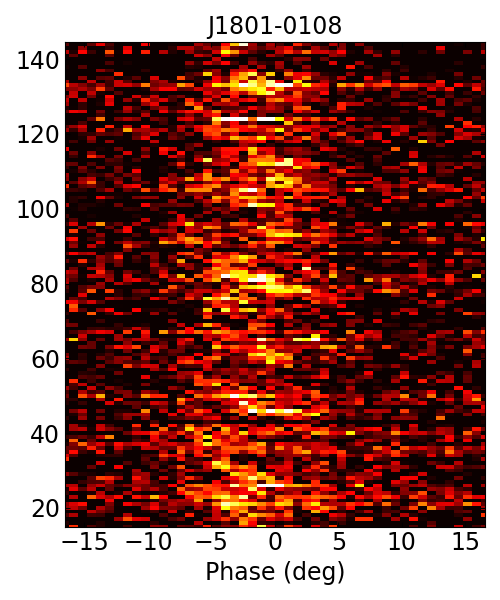}
  \includegraphics[width=4cm,height=5.5cm]{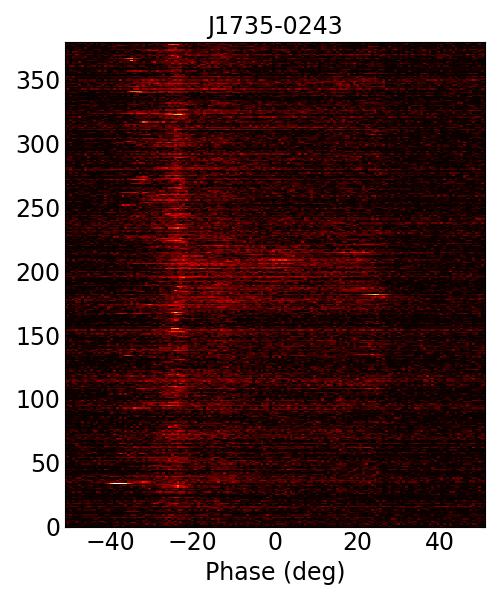}
  \caption{The single pulse stacks of four pulsars exhibiting unique single-pulse phenomena are presented, with the first three being newly discovered and the fourth being a known pulsar. All display either mode switching or subpulse drifting behavior.
    \label{fig_singlepulse}}
\end{figure*}

By examining the single pulse stacks of the new and known pulsars from this survey, we observed that five of the pulsars display mode changing or subpulse drifting behavior during the observations. The single pulse stacks of these pulsars are shown in Figure\,\ref{fig_singlepulse}.

For J1745$-$0059, a distinct intensity modulation is observed in its second component, which is not evident in the first component. This suggests two distinct emission modes, distinguished by the intensity of the second component.
J1800$-$0059 and J1801$-$0108 both exhibit subpulse drifting behavior, with the distinction that their subpulse drift trajectories are in opposite directions.
For J1735$-$0243, a known pulsar discovered by \citet{2013ApJ...763...80Boyles} using the Green Bank Telescope at 350~MHz, our five-minute observations reveal a mode change at around pulse number 200. In the normal mode, the first component is more intense relative to others, while in the anomalous mode, the intensity of each component is comparable. Notably, in our observations, the anomalous pattern accounts for only about one-tenth of the total observation time. 
While we observed single-pulse phenomena in these pulsars over a short period, detailed and statistical characterization of their single pulses will require longer and more sensitive observations.

%%%%%%%%%%%%%%%%%%%%%%%%%%%%%%%%%%%%%%%%%%
%%%%%%%%%%%%%%%%%%%%%%%%%%%%%%%%%%%%%%%%%%

\section{Discussion and conclusion}\label{sec_4}

This study reports the results of a high Galactic latitude survey conducted with FAST. Within a survey area of approximately 15 square degrees, we discovered 15 new pulsars, comprising 11 normal pulsars and four MSPs. Among the MSPs, one is isolated, while the other three are in binary systems. 
Notably, these are the first Galactic field MSPs discovered in this region.
For the eleven newly discovered normal pulsars, follow-up timing observations with Murriyang or FAST allowed us to successfully obtain coherent timing solutions for three pulsars: J1745$-$0059, J1746$-$0156, and J1800$-$0059. Furthermore, we analyzed the single pulses of both new and known pulsars observed in this survey and found that four pulsars exhibited interesting single pulse phenomena. J1800$-$0059 and J1801$-$0108 display subpulse drift behavior with opposite drift trajectories, while J1745$-$0059 and J1735$-$0243 exhibit two distinct emission modes. Detailed and statistical properties of these single pulses require longer and highly sensitive observations.

Three of our discoveries are MSPs in binary systems. The orbital period ($P_{\rm b}$) and companion mass ($M_{\rm c}$) of PSRs~J1748$-$0224 and J1801$-$0144 are consistent with the $P_{\rm b}$--$M_{\rm c}$ relation for low-mass white dwarf (WD)--MSP binary systems~\citep[e.g.,][]{ts99}, suggesting that their companions are likely helium (HE) WDs. The companion of PSR~J1742$-$0237 has a mass of 0.46\,M$_\odot$ (at an inclination angle of 60 degrees), significantly more massive than the prediction for a HE-MSP binary with an orbital period of 12.5\,days. Therefore, PSR~J1742$-$0237 could potentially be in a binary system with a carbon-oxygen WD~\citep[e.g.,][]{hwh+18}. Optical identification and/or precise mass measurement of MSP companions are required to confirm the nature of these systems.

\begin{figure}
  \centering 
  \includegraphics[width=0.5\textwidth]{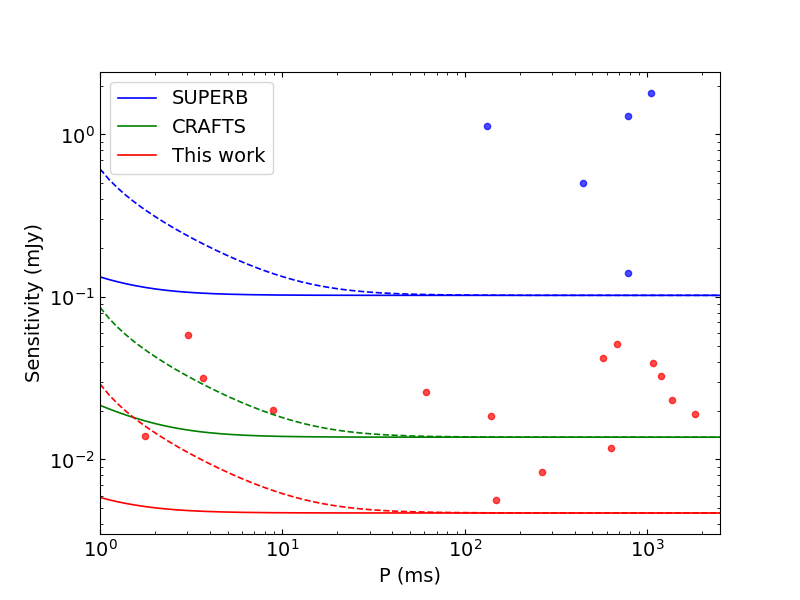} 
  \caption{ The sensitivity of the three surveys (SUPERB, CRAFTS and this work) as a function of pulsar spin period for different DM values, assuming a 10\% duty cycle. The solid and dashed lines represent ${\rm DM}=50$\,pc\,cm$^{-3}$ and 300\,pc\,cm$^{-3}$, respectively. Estimated flux densities of our discoveries are shown as red dots. Flux densities of known pulsars in the surveyed region were obtained from the ATNF pulsar catalogue and shown as blue dots.
    \label{fig_sensitivity}}
\end{figure}

We estimated the sensitivity of our survey to MSPs using the radiometer equation~\citep{2004hpa..book.....L}. We used a gain of $G=16$\,K/Jy for FAST and observing parameters described in Section \ref{sec_2}. Our sensitivity was compared with previous surveys at high Galactic Latitudes, including SUPERB~\citep{2018MNRAS.473..116Keane} and CRAFTS~\citep{2019SCPMA..6259506ZhangK}. In Fig.\,\ref{fig_sensitivity} we show sensitivities as a function of the spin period for different DMs, assuming a 10\% duty cycle. The sensitivity of our survey is more than one order of magnitude higher than that of SUPERB. Compared to CRAFTS, a drift-scan survey, our survey is approximately three times more sensitive, primarily due to the longer integration time per pointing.
In addition, we estimated flux densities of newly discovered pulsars using profiles in Fig.~\ref{fig_allprofiles} and the radiometer equation~\citep{2004hpa..book.....L} and showed them as red dots in Fig.~\ref{fig_sensitivity}. For comparison, the flux densities of known pulsars in our survey region, obtained from the ATNF pulsar catalogue~\citep{2005AJ....129.1993M}\footnote{\url{http://www.atnf.csiro.au/research/pulsar/psrcat.}}, are shown as blue dots. Our newly discovered pulsars exhibit significantly lower flux densities, further highlighting FAST’s capability to detect faint pulsars.

The existence of a population of MSPs within the Galactic Bulge remains an open question. If the GCE originates from a population of MSPs, several studies~ \citep[e.g.,][]{cdd+16,bcc+21} have shown that their spatial distribution could extend to Galactic latitudes as high as $\sim20$\degree and differs from that of the Galactic disk population. Furthermore, \citet{cdd+16} demonstrated that the absence of detected MSPs in the Bulge is primarily due to the limited sensitivity of pulsar searches in these regions. These findings highlight the importance of deep pulsar surveys at high Galactic latitudes near the Bulge.

To determine whether any of our detected MSPs are associated with the Galactic Bulge population, we compared our findings with simulations of both disk and Bulge MSPs. For the Galactic disk population, we employed the \textit{PSRPOPPY} software \citep{2014MNRAS.439.2893Bates}, following the procedure described in \citet{2017MNRAS.472.1458DaiS}. Our simulations incorporated spin period, luminosity, and spectral distributions based on previous studies \citep[e.g.,][]{2004A&A...422..545Yusifov, 2006ApJ...643..332Faucher, 2006MNRAS.372..777Lorimer, 2013MNRAS.434.1387Levin, 2015MNRAS.450.2185Lorimer}. 
For the Bulge MSP population, we adopted a model based on the distribution of red clump giants \citep{2013MNRAS.434..595C}, informed by recent $\gamma$-ray results \citep{2018NatAs...2..819B}. This model was implemented in \cite{2021PhRvD.104d3007B} with an updated Sun–Galactic center distance of\,8.3 kpc. The other characteristics of the Bulge population are similar to those of the disk population.
Our simulations predicted that approximately nine disk MSPs should be detectable in our survey, whereas no Bulge MSPs are expected to be detected (average over 100 simulations). Based on these results, we conclude that the MSPs discovered in this survey are most likely located in the Galactic disk. 

\begin{figure}
  \centering \includegraphics[width=0.45\textwidth]{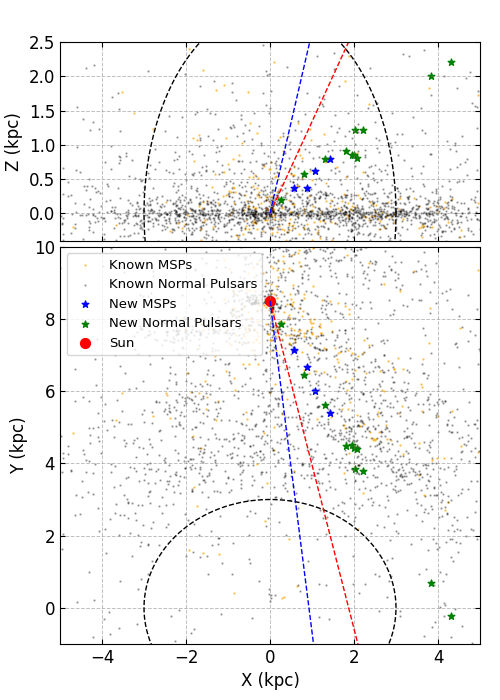}   
  \caption{Galactic distribution of 15 pulsars discovered in this survey (green: normal pulsars; blue: MSPs). Distances were estimated using the YMW16 electron density model~\citep{YMW16}. Known normal pulsars and MSPs are indicated by gray and orange dots, respectively. We only included known pulsars with a distance from the Galactic plane ($|Z|$) less than 2.5\,kpc. The black dashed line shows a distance of 3\,kpc from the Galactic Center, which is indicative of the GCE extension~\citep{2015PhRvD..91f3003C}. The blue and red dashed lines indicate the innermost boundaries of a similar survey if the FAST snapshot mode could extend to declinations of  $-15$\degree and $-10$\degree, respectively.
    \label{fig_position}}
\end{figure}

Due to FAST's snapshot mode declination limit, our survey only covers a region near the Bulge's edge (see Fig.~\ref{fig_position}). However, FAST's full observing range extends to $-14\degree$, so observing at declinations of $-15\degree$ or even $-10\degree$ would allow us to probe further into the Bulge. To evaluate the feasibility of detecting Bulge MSPs in future FAST observations, we estimated the number of detectable Bulge MSPs for a survey similar to ours but with an extended declination limit. 
Using the same methodology employed to estimate the number of Bulge MSPs in our current observations, we predict FAST would detect approximately $\sim1$ (or 3) Bulge MSPs if it observed as far south as 10$\degree$ (or 15$\degree$). This low detection rate is likely due to the angle between the Galactic Bar's major axis and the Sun-Galactic center line~\citep[e.g.,][]{2016ARA&A..54..529B}, resulting in fewer Bulge MSPs visible in the FAST sky. Our simulations indicate that deep surveys of the Galactic Bulge from the southern hemisphere have a greater potential for discovering Bulge MSPs, as they provide direct observations through the Galactic Bar. However, large-scale FAST surveys at high Galactic latitudes remain valuable for studying the spatial distribution of disk MSPs and probing the Bulge MSP population, given the current gaps in our understanding.

%%%%%%%%%%%%%%%%%%%%%%%%%%%%%%%%%%%%%%%%%%

\acknowledgments

This work made use of the data from FAST (Five-hundred-meter Aperture Spherical radio Telescope, \url{https://cstr.cn/31116.02.FAST}). FAST is a Chinese national mega-science facility, operated by National Astronomical Observatories, Chinese Academy of Sciences.
Murriyang, the Parkes radio telescope, is part of the Australia Tele-
scope National Facility (\url{https://ror.org/05qajvd42}) which is funded
by the Australian Government for operation as a National Facility
managed by CSIRO.
This work is supported by the National Natural Science Foundation of China (Nos. 12273008, 11988101, 12041303, 12041304), the National SKA Program of China (Nos.2022SKA0130100, 2022SKA0130104), the Natural Science and Technology Foundation of Guizhou Province (No. [2023]024), the Foundation of Guizhou Provincial Education Department (No. KY (2020) 003).

\bibliography{ref}{}

\bibliographystyle{aasjournal}

\end{document}